\documentclass[12pt]{iopart} 
\usepackage{iopams}  
\usepackage{graphicx}
\usepackage{dcolumn}
\usepackage{bm}
\usepackage{hyperref}

\begin{document}
\title{Principles of an atomtronic transistor}

\author{Seth C. Caliga$^1$, Cameron J. E. Straatsma$^2$, Alex A. Zozulya$^3$, and Dana Z. Anderson$^1$}
\address{$^1$ Department of Physics, University of Colorado, and JILA, University of Colorado and National Institute for Standards and Technology, Boulder CO 80309-0440}
\address{$^2$ Department of Electrical, Computer, and Energy Engineering, University of Colorado, and JILA, University of Colorado, Boulder CO 80309-0440}
\address{$^3$Department of Physics, Worcester Polytechnic Institute, Worcester MA 01609-2280}
\address{Department of Physics, University of Colorado, and JILA, University of Colorado and National Institute for Standards and Technology, Boulder CO 80309-0440}

\begin{abstract}
A semiclassical formalism is used to investigate the transistor-like behavior of ultracold atoms in a triple-well potential. Atom current flows from the source well, held at fixed chemical potential and temperature, into an empty drain well. In steady-state, the gate well located between the source and drain is shown to acquire a well-defined chemical potential and temperature, which are controlled by the relative height of the barriers separating the three wells. It is shown that the gate chemical potential can exceed that of the source and have a lower temperature. In electronics terminology, the source-gate junction can be reverse-biased. As a result, the device exhibits regimes of negative resistance and transresistance, indicating the presence of gain. Given an external current input to the gate, transistor-like behavior is characterized both in terms of the  current gain, which can be greater than unity, and the power output of the device.   
\end{abstract}
\maketitle

\section{\label{sec:Intro}Introduction\protect\\}
Moore's law observes that the number of transistors on an integrated circuit doubles approximately every two years.  Quite apart from a comment about the remarkable evolution of  electronics, Moore's law highlights the cardinal role of the transistor in modern technology.  What is it that distinguishes the transistor from other electronic elements?  The canonical transistor is a three-terminal device that derives its utility by enabling a small electric current or voltage to control a large current.  In particular, transistors are active devices that provide electrical gain---gain that can be used to amplify a signal, sustain coherent oscillation, perform switching, provide storage, and implement logic. Such are a few of the many functions that reflect the ubiquity of the transistor in modern electronics.  

Transistor-like behavior has been studied in both cold and ultracold atomic systems revealing a variety of approaches for realizing an atomtronic transistor.  The mechanisms responsible for transistor action explored thus far can be distinguished into three categories: (1) interatomic interactions of a single species in a single state, $A$, i.e., $A_\mathrm{in}$ controls $A_\mathrm{out}$~\cite{Seaman:2007kx, Stickney:2007ix, Pepino:2009jb, Pepino:2010}; (2) Interatomic interactions of two species or states, $A$ and $B$, i.e., $B_\mathrm{in}$ controls $A_\mathrm{out}$~\cite{Gajdacz:2014, Micheli:2004}; and, (3) an external coupling field, $C$, that drives internal state transitions of atom $A$, i.e., $C_\mathrm{in}$ controls $A_\mathrm{out}$~\cite{Vaishnav:2008, Benseny:2010}. The atomtronic transistor described in this work falls into the first category and follows from the analogy with electronic transistors, which operate based on the flow or buildup of a single carrier species, namely electrons.  The first two categories, both of which derive functionality from interatomic interactions, are differentiated solely by experimental implementation. By introducing a second species or state, $B$, to control the flow of $A$, the transistor system becomes more complex, requiring the superposition of two species- or state-dependent trapping potentials.  By contrast, the third category utilizes a coupling, $C$, other than atoms to control the flow of $A$. In such systems the coupling strength of field $C$ is converted into the flow of atoms $A$, a process akin to transduction.  As in electronics, the development of various types of transistors for specific applications emphasizes their utility.

Driving an interest in atomtronics is the possibility of developing a paradigm for addressing problems in quantum signal and information processing that parallels the power of electronics in the classical realm. Thus, an atomtronic version of transistor action should be a central theme of research in atomtronic devices and circuits. Utilizing ultracold atoms in place of  electrons,  atomtronics lives in a challenging arena of physics involving many-body, interacting, and open quantum systems in non-thermal equilibrium.   It is likely not possible to develop complete descriptions of any but the simplest atomtronic circuits.  Indeed, most experiments thus far involve fully quantum atomtronic circuits, i.e., operating as a closed system that is described by unitary evolution~\cite{Moulder:2012, Ryu:2013, Eckel:2014, Ryu:2015}.  Transistor-like gain, on the other hand, requires an open system approach~\cite{Seaman:2007kx, Stickney:2007ix, Pepino:2009jb, Pepino:2010, Gajdacz:2014, Chow:2015}.  Moreover, accepting the connection between information processing and entropy change suggests that dissipation and heat generation are fundamental aspects to be taken into account in atomtronic circuit design~\cite{Landauer:1961, Rothstein:1951, Miller:1959,  Jendrzejewski:2014, Faist:2015}. In classical circuits, electrons are strongly coupled to the thermal environment of the medium in which they propagate.   Ultracold atomic systems, by contrast, are isolated from the surrounding environment such that dissipation and heat generation contribute to the state of the system. To make headway toward the understanding of atomtronic circuits one can either treat such dissipation as a perturbation to an otherwise quantum system, or introduce quantum effects to an otherwise classical system. Both can be instructive. It is interesting to note that the majority of  electronic devices and circuit behavior can be understood within the framework of thermodynamics, including, for example, transistor action.  

This paper presents a semiclassical analysis of the flow of ultracold atoms having finite temperature, $T$, and chemical potential, $\mu$, within a triple-well potential designed to mimic the behavior of an electronic transistor.   Here, each barrier that separates two adjacent wells is analogous to the built-in potential in the space charge region of a semiconductor junction or the work function of the cathode in a vacuum tube~\cite{Sze:1969, Smith:1971cd}.  Much like the thermionic emission in these electronics examples~\cite{Bethe:1942}, atom currents in our system flow due to those atoms that are energetic enough to traverse the repulsive barrier. Thus, atom currents are described using a formalism common to evaporative cooling~\cite{Luiten:1996ja, Walraven:1996qd}. Building upon these previous works, we define relations for atomic currents driven by thermal and chemical potentials. Non-equilibrium dynamics of double-well ultracold atomic systems have been studied, exemplifying the role that gradients in temperature and chemical potential play in driving the system towards thermal equilibrium~\cite{Shin:2004, Brantut:2012mc, Brantut:2013he, Hazlett:2013, Labouvie:2015nd}. 

The name ``transistor," an amalgam of ``transfer" and ``resistor," refers to the effective resistance, or transresistance of the device when treated as a two-port, input/output device. It is precisely that this resistance can be negative that corresponds to the gain that transistors are known for~\cite{Davis:2011cd, Edson:1953vt}. More than just a conceptual convenience, the negative resistance of a transistor is a thermodynamic reality: the transresistance corresponds to negative power dissipation during current flow~\cite{Willardson:1971ss}.  Here, the concepts of negative transresistance and gain are applied to the atomic system to investigate the principles of an atomtronic transistor where transistor-like behavior arises due to the interaction of currents between three wells.  

In section~\ref{sec:Model}, a steady-state model is presented that determines the chemical potential and temperature within each well.  Controlling the differences in chemical potential and temperature of the populations in two adjacent wells can be understood as the analogue to biasing an electronic transistor to the desired quiescent point (Q-point).  We begin by describing the trapping potential that forms the three wells of the transistor.  The Q-point is shown to be controlled by the relative height of the barriers that separate the three wells, and a reverse-biased regime is revealed in which the chemical potential of the ensemble in the gate is larger than that of the source.  The resulting temperature and chemical potential differences are addressed in relation to the effective static resistance of the source-gate junction.

Transistor-like functionality of the device is established in section~\ref{sec:Operation}, where current gain and maximum power output are calculated given an externally applied input current to the gate well.  Here, transistor action arises due to the interplay between the input current and the steady-state thermodynamic properties of the gate.  Depending on the Q-point, the current into the drain is either attenuated or amplified in response to the applied gate current, demonstrating a current gain greater than unity.  Additionally, the transresistance is found to be negative indicating the active nature of the transistor in that it draws power from a reservoir to control power output to a load. The maximum power to an impedance matched load is then shown for a range of Q-points.

Finally, section~\ref{sec:Expt} discusses relevant characteristics of the trapping potential in realizing results from~\ref{sec:Model} and~\ref{sec:Operation}.  Key parameters include the trap frequencies and configuration of the barriers that separate the triple-well potential. Explicit parameter values are given for an experiment that demonstrates quasi-steady-state control over the chemical potential and temperature throughout a triple-well system akin to the results in section~\ref{sec:Model}~\cite{Caliga:2015}.

\section{\label{sec:Model}Steady-state model}
\subsection{The trapping potential}
\begin{figure}
\begin{center}
\includegraphics[width=3.25in]{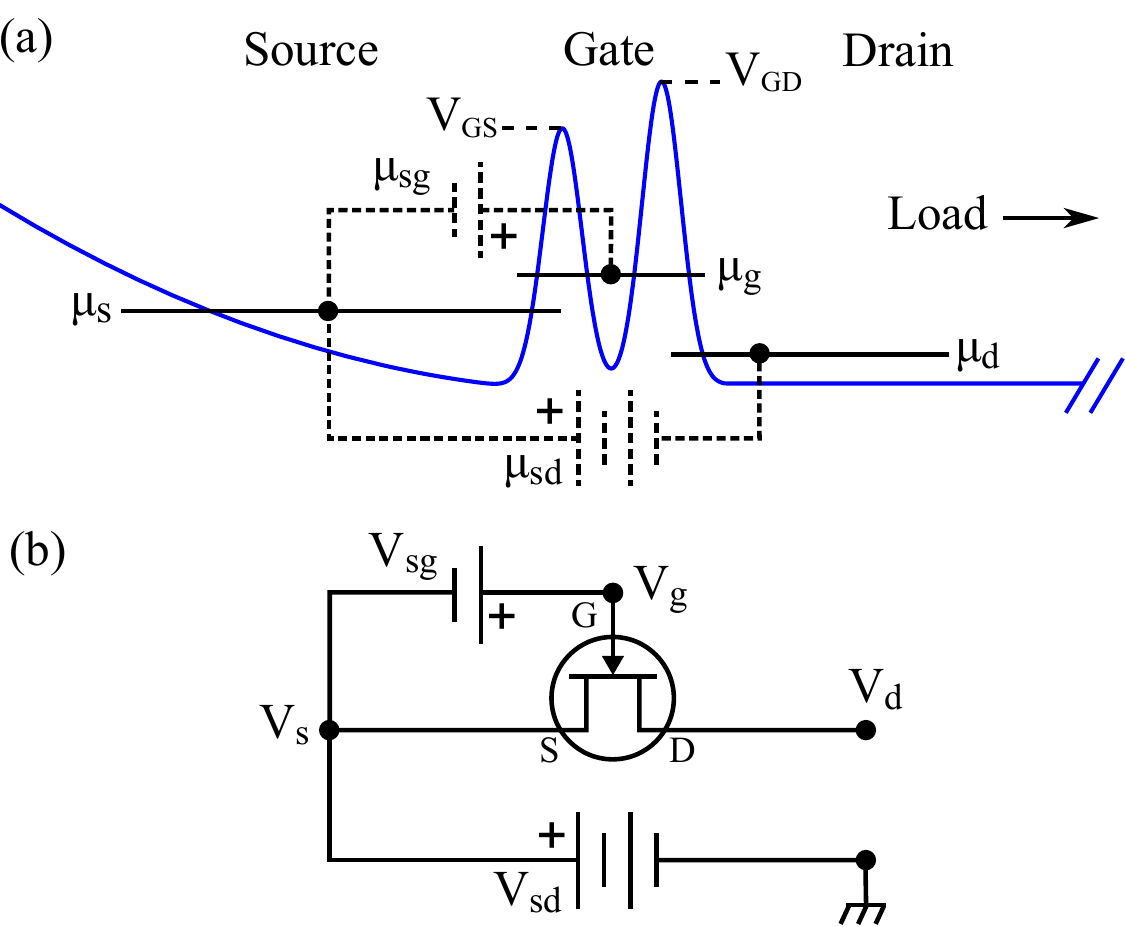}
\caption{ \label{fig:Potential} (a) Triple-well potential with source, gate, and drain wells.  Chemical potential levels in each well are depicted in the context of potential biases. The current output into the drain can then be coupled to the desired `load.' In the model presented here, the flow of atoms through the potential is controlled by adjusting the source-gate chemical potential difference via an external current input to the gate .  (b) A simplified schematic of an electronic common-source amplifier circuit with analogous functionality to its atom based counterpart.}
\end{center}
\end{figure}

Our semiclassical treatment of an atomtronic transistor begins by borrowing the nomenclature of the semiconductor field-effect transistor.  The three regions of the triple-well potential, shown in Fig.~\ref{fig:Potential}(a), are labeled the ``Source," ``Gate," and ``Drain" wells. Emulating an experimentally viable system we take the trapping potential to be cigar shaped with corresponding harmonic trap frequencies $\omega_x \ll \omega_{\perp}$. This potential is sectioned in the longitudinal direction by two repulsive Gaussian barriers having peak heights $V_{\mathrm{GS}}$ and $V_{\mathrm{GD}}$, respectively, to form the source, gate, and drain wells. The separation and width of the barriers determines the longitudinal gate well trap frequency. Furthermore, the longitudinal trap axis of the source well is assumed to be half harmonic for simplicity. The degree of overlap of the two barriers contributes to a potential bias in the gate, $V_\mathrm{G,0}$, with respect to the source and drain.  Finally, the drain well is modeled as a reflectionless output port that would feed into a subsequent circuit element.  

\subsection{Current flow mechanism}
Previous works have modeled the flow of atoms across a potential barrier in order to study both the selective removal of atoms involved in evaporative cooling as well as the reverse process in trap loading~\cite{Luiten:1996ja,Walraven:1996qd,Roos:2003tp}. Here, the expressions in Ref.~\cite{Roos:2003tp} are modified to include the chemical potential due to the presence of a Bose-Einstein condensate (BEC) in each well, the analogue of an applied bias voltage.  Thus, the atom currents in the triple-well system are defined by the set of equations:
\begin{eqnarray}\label{eqn:currents}
I_{\mathrm{sg}} & = & \gamma_{\mathrm{s}} N_{\mathrm{th,s}}\mathrm{exp}[-(V_{\mathrm{GS}}-\mu_{\mathrm{s}})/kT_{\mathrm{s}}], \nonumber \\
I_{\mathrm{gs}} & = & \gamma_{\mathrm{g}} N_{\mathrm{th,g}}\mathrm{exp}[-(V_{\mathrm{GS}}-V_\mathrm{G,0}-\mu_{\mathrm{g}})/kT_{\mathrm{g}}], \nonumber \\
I_{\mathrm{gd}} & = & \gamma_{\mathrm{g}} N_{\mathrm{th,g}}\mathrm{exp}[-(V_{\mathrm{GD}}-V_\mathrm{G,0}-\mu_{\mathrm{g}})/kT_{\mathrm{g}}],
\end{eqnarray}
where k is the Boltzmann constant, the product $\gamma_{i}N_{\mathrm{th,}i}$ is the effective collision rate of thermally excited atoms in the $i$-th well, and the exponential factor reflects the thermodynamic probability that an atom possesses sufficient energy to traverse the barrier.  Note that the subscript order indicates the direction of current flow, e.g. $I_{\mathrm{sg}}$ describes the atom flux from the source to the gate well, and capitalized subscripts indicate model parameters. 

We consider the range of temperatures below the critical temperature, $T_\mathrm{c}$, but above the temperature associated with the interaction energy per particle, $T_\mathrm{0}$.  In this regime, $T_\mathrm{c} \gtrsim T > T_\mathrm{0}$, the elastic collision rate responsible for the currents is dominated by collisions between thermally excited atoms~\cite{Pethick:2002tn}.  The equilibrium collision rate ($\gamma = \sqrt{2/\pi}n_{th}\sigma_0\Delta\mathrm{v}$) then depends only on the peak density of the thermal component, $n_{th}=\zeta(3/2)/\lambda_{th}^3$, where $\sigma_\mathrm{0}$ is the collision cross section, $\Delta\mathrm{v}$ is the mean thermal velocity, $\zeta(z)$ is the Riemann zeta function, and $\lambda_{th}$ is the thermal DeBroglie wavelength~\cite{Pethick:2002tn}. Furthermore, scattering at the energies considered is purely s-wave. Thus, the collision rate for the $i$-th well is given by
\begin{equation}\label{eqn:gamma}
\gamma_{i}=32\pi^{2}\zeta(3/2)m(a_{s}kT_{i})^{2}/h^{3}, 
\end{equation}
where $m$ is the atomic mass and $a_s$ is the s-wave scattering length.

\subsection{Source and gate well ensembles in steady-state}
Steady-state circuit operation is analyzed by enforcing particle number and energy conservation, expressed using analogues of Kirchhoff's current and voltage laws:
\begin{eqnarray}
I_{\mathrm{sg}} = I_{\mathrm{gs}}+I_{\mathrm{gd}}, \label{eqn:Kirchoff} \\
I_{\mathrm{sg}}(V_{\mathrm{GS}}+\kappa_{\mathrm{GS}}kT_{\mathrm{s}}) 
= I_{\mathrm{gs}}(V_{\mathrm{GS}}+\kappa_{\mathrm{GS}}kT_{\mathrm{g}})+I_{\mathrm{gd}}(V_{\mathrm{GD}}+\kappa_{\mathrm{GD}}kT_{\mathrm{g}}), \label{eqn:Energy}
\end{eqnarray}
\begin{figure}[floatfix]
\begin{center}
\includegraphics[width=4in]{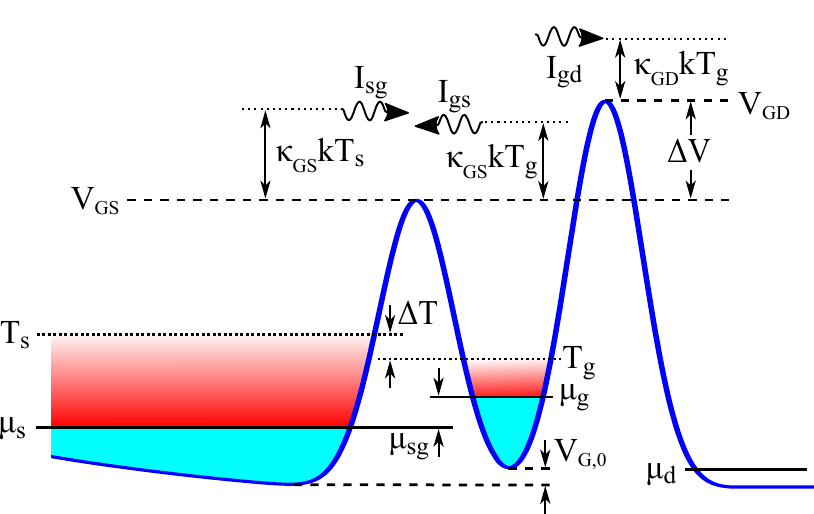}
\caption{\label{fig:GateDynamics} Schematic of atom and energy flow across the gate barriers.  The atom current carries with it an energy in excess of the barrier height by the factor $\kappa k T$ relative to each barrier.  Light blue shading indicates the offset in energy of the thermal component, shown in red, due to the condensate chemical potential.  An example steady-state for source and gate ensembles is shown, in which the temperature drop, $\tau$, is positive and the chemical potential drop, $\mu_\mathrm{sg}$, is negative.}
\end{center}
\end{figure}%
where the $\kappa$'s, which indicate the average excess energy of atoms traversing the barriers, are of order unity~\cite{Roos:2003tp}.  It is supposed that some external reservoir supplies atoms to the source well, maintaining a fixed chemical potential, $\mu_{s}$, and temperature, $T_{s}$.  For barrier heights that are large compared to the chemical potential and thermal energies, the three wells are weakly coupled by thermal atoms that have sufficient energy to traverse the barriers. Thus, we assume that the gate acquires a well-defined chemical potential, $\mu_\mathrm{g}$, and temperature, $T_\mathrm{g}$, once a steady-state is reached. Additionally, we impose that current into the drain is removed from the system such that no current flows from the drain back towards the gate. An illustration of the model system is shown in Fig.~\ref{fig:GateDynamics}.

With the source well ensemble held constant, we seek steady-state values of the gate chemical potential and temperature in terms of $\mu_\mathrm{s}$ and $T_\mathrm{s}$.  It proves useful to define a temperature drop $\tau$ and barrier height difference $v$, both normalized to the source temperature:
\begin{eqnarray}\label{eqn:DeltaT}
\tau & \equiv & \left( {{T_{\mathrm{s}}} - {T_{\mathrm{g}}}} \right)/{T_{\mathrm{s}}}\equiv\Delta T/T_{\mathrm{s}}, \\
v & \equiv & (V_{\mathrm{GD}}-V_{\mathrm{GS}})/kT_{\mathrm{s}}\equiv\Delta V/kT_{\mathrm{s}}.
\end{eqnarray}
The value $v$ is referred  to as the feedback parameter.  Using the currents defined in Eq.~(\ref{eqn:currents}), along with Eqs.~(\ref{eqn:Kirchoff}) and~(\ref{eqn:Energy}), it is possible to derive relations for both the temperature drop,
\begin{equation}\label{eqn:DeltaTrans}
    \tau  = {e^{ - \upsilon/\left( {1 - \tau } \right)}}\frac{{\upsilon + \left( {{\kappa_{\mathrm{GD}}} - {\kappa_{\mathrm{GS}}}} \right)}}{{{\kappa_{\mathrm{GS}}} + {\kappa_{\mathrm{GD}}}{e^{ - \upsilon/\left( {1 - \tau } \right)}}}},
\end{equation}
and the normalized chemical potential drop,
\begin{eqnarray}\label{eqn:PotentialDrop}
\hat{\mu }_\mathrm{sg} & = & \left(1-\tau \right)\ln\left[\left(1-\tau\right)^{4}\left(1+e^{-\frac{\upsilon}{1-\tau }}\right)\right] - \tau\left(\hat{V}_{\mathrm{GS}}-\hat{\mu }_{\mathrm{s}}\right) \nonumber \\
& + &(1-\tau )\ln \left[\frac{1-\tau+\frac{\zeta(2)}{\zeta(3)} \left(\hat{\mu }_{\mathrm{s}}-\hat{V}_{\mathrm{G,0}}-\hat{\mu }_\mathrm{sg}\right)}{1+\frac{\zeta(2)}{\zeta(3)} \hat{\mu }_{\mathrm{s}}}\right],
\end{eqnarray}
where $\mu_\mathrm{sg} \equiv \mu_\mathrm{s}-\mu_\mathrm{g}$ and the hat ($\hat{~}$) indicates quantities normalized to the source temperature.  Within these equations, $e^{-\upsilon/(1-\tau)}$ represents the ratio of forward current into the drain relative to current fed back into the source, $I_\mathrm{gd}/I_\mathrm{gs}$.  In deriving Eq.~(\ref{eqn:PotentialDrop}) we have used the fact that the thermal and chemical potential drops are zero at $\upsilon = \infty$ along with Ref.~\cite{Pethick:2002tn} to determine the ratio of $\gamma_{\mathrm{g}} N_{\mathrm{th,g}}/\gamma_{\mathrm{s}} N_{\mathrm{th,s}}$ that satisfies Eq.~(\ref{eqn:Kirchoff}).  In the absence of an external gate input, Eq.~(\ref{eqn:PotentialDrop}) shows that the souce-gate junction is self-biased to some Q-point characterized by $\mu_\mathrm{sg}$. The magnitude of this bias is primarily dependent on the feedback parameter, both directly and through the temperature drop.  This bias is illustrated in Fig.~\ref{fig:Potential}(a) within the atomic system and Fig.~\ref{fig:Potential}(b) for an equivalent, simplified electronic circuit.  Equations~(\ref{eqn:DeltaTrans}) and~(\ref{eqn:PotentialDrop}) form the basis of our model and can be solved self-consistently in order to characterize the thermodynamic variables of the triple-well system in steady-state. 

To calculate the values of $\tau$ and $\hat{\mu}_\mathrm{sg}$, the $\kappa$ parameters of the trapping potential must be determined.  Trap geometry has a strong effect on the average energy removed by an atom leaving via a controlled trajectory. For atoms escaping isotropically from a potential well, as in typical evaporative cooling schemes, $\kappa \simeq 1$~\cite{Luiten:1996ja,Walraven:1996qd}.  However, limiting the escape trajectory to purely 1D, for instance along the loose axis of a cigar-shaped trap, $\kappa \simeq 2.9$ for truncation parameters $\eta \equiv V/T = 4 \-- 7$~\cite{Roos:2003tp}. The difference in $\kappa$ factors is a direct result of allowed escape trajectories. Molecular dynamics simulations akin to those used in Refs.~\cite{Roos:2003tp,Wu:1997dm} were performed to confirm $\kappa \simeq 2.9$ for the geometry and feedback parameters modeled here. 

\begin{figure}[floatfix]
\begin{center}
\includegraphics[width=5.25in]{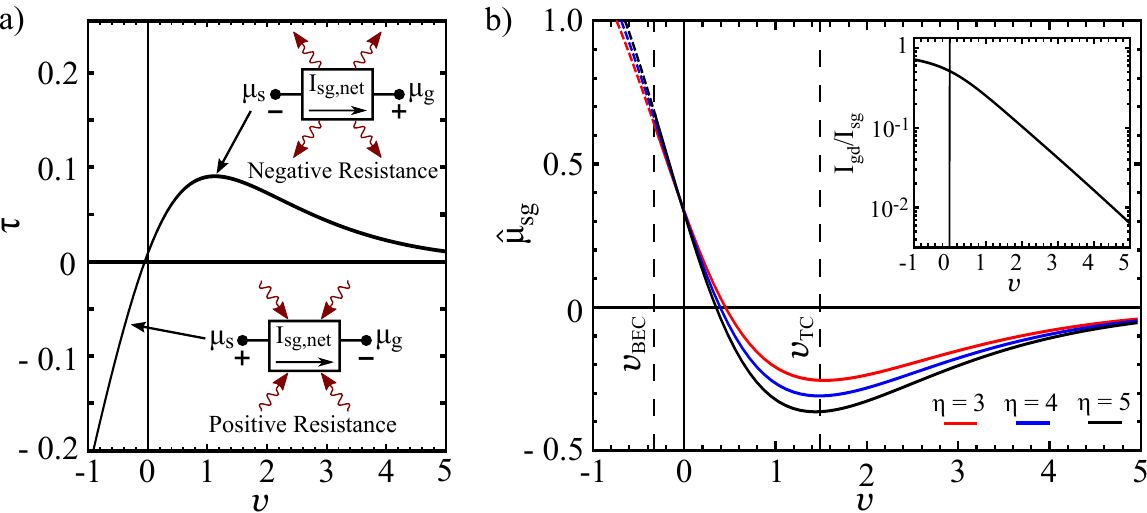}
\caption{\label{fig:Temperature_Drop} Quiescent point characteristics. (a) Plot of the temperature drop vs.~feedback parameter.  Insets illustrate power dissipation due to the static source-gate resistance.  Negative $\upsilon$ results in a negative temperature drop in the direction of net current across the source-gate barrier, indicating heat transfer into the gate and $T_\mathrm{g}>T_\mathrm{s}$ in steady-state. Conversely, positive $\upsilon$ results in net cooling (negative power dissipated) and $\tau$ increases until $\mu_\mathrm{sg}$ peaks. As $v\rightarrow\infty$, $I_\mathrm{gd}\rightarrow 0$ and equilibrium between the gate and source wells is reached. (b) Plot of the chemical potential drop vs.~feedback parameter.  Multiple curves show the dependence on the truncation parameter, $\eta$. Threshold feedback parameters are shown for the onset of BEC ($\upsilon_{\mathrm{BEC}}$) in the gate and negative transconductance ($\upsilon_{\mathrm{TC}}$). The inset shows the steady-state current output of the transistor, normalized to the source-gate current, as a function of $\upsilon$.}
\end{center}
\end{figure}

The temperature and chemical potential drops are shown as functions of the feedback parameter in Fig.~\ref{fig:Temperature_Drop}(a), with $\kappa_{\mathrm{GS}}=\kappa_{\mathrm{GD}}=2.9$. The behavior shown by the temperature drop is somewhat non-intuitive in that for values of positive feedback ($V_\mathrm{GD}>V_\mathrm{GS}$), the temperature of the gate is actually {\em{lower}} than the source, despite current from the source into the gate having an average energy $V_{\mathrm{GS}}+\kappa_{\mathrm{GS}}kT_\mathrm{s}$.  Furthermore, Fig.~\ref{fig:Temperature_Drop}(b) shows that there will generally exist a range of feedback parameters for which the chemical potential drop is negative, meaning the source-gate junction is reverse-biased.  The inset of Fig.~\ref{fig:Temperature_Drop}(b) shows the gate-drain current normalized to the source-gate current versus the feedback parameter.  The ratio of currents is independent of $\eta$, but the absolute magnitude of the $I_\mathrm{gd}$ depends strongly on the source ensemble and $\eta$ given the exponential nature of the currents in Eqs.~(\ref{eqn:currents}).

Also shown in Fig.~\ref{fig:Temperature_Drop}(b) are two threshold feedback parameters.  The first, $\upsilon_\mathrm{BEC}$, indicates the formation of a BEC in the gate well.  For $\upsilon\geq\upsilon_\mathrm{BEC}$ the balance of particle and energy currents leads to a steady-state gate well ensemble with sufficiently high phase-space density to condense. The second threshold, $\upsilon_\mathrm{TC}$, indicates the feedback parameter above which negative transconductance occurs for the steady-state parameters.  Section~\ref{sec:ACOp} provides an in-depth discussion of this threshold.

To better understand the steady-state results, one can consider the power dissipated within the gate well of the transistor, $P_\mathrm{sg} = I^2_\mathrm{sg,net} R_\mathrm{sg}$. Here $I_\mathrm{sg,net}=I_\mathrm{sg}-I_\mathrm{gs}>0$ is the total atom current flowing into the gate. Note that the power dissipated within the gate well does not include the gate to drain current, as power output to the drain is available to do work on a connected load. The static source-gate resistance in steady-state is
\begin{eqnarray}\label{resistance}
R_{\mathrm{sg}} & \equiv & \mu_{\mathrm{sg}}/I_{\mathrm{sg,net}}.
\end{eqnarray}
Given the sign of the chemical potential drop, $R_{\mathrm{sg}}$ can be negative indicating an ohmic cooling synonymous with the evaporative cooling process, which leads to positive $\tau$.

\section{\label{sec:Operation} Device operation}
Using the steady-state results from section~\ref{sec:Model}, we  investigate transistor action in the triple-well system. The interaction between an external current input to the gate and the steady-state gate population is shown to modify the gate ensemble and therefore the current into the drain.  The change in drain current relative to the current injected into the gate yields information about the current gain.  The transconductance, inverse of the transresistance, is also calculated in order to determine the power dissipated at the output of the transistor.  As alluded to previously, the sign of the transconductance determines the sign of power dissipation.  Negative power dissipation reflects the ability of the device to supply power.
\subsection{Current gain}
Within the steady-state model presented, current gain is the most readily studied quantity since voltage and power gain require more explicit knowledge of the connected `load.'  To study gain, we use a method similar to determining the two-port admittance parameters of an electronic circuit, which is commonly used to describe transistor action in electronic transistors~\cite{Bardeen:1949}.  The coupled system of equations for the triple-well atomic system is given by
\begin{equation}\label{eqn:Admittance}
\left( \begin{array}{rl}
\mathrm{d}I_{\mathrm{N,g}}\\	
\mathrm{d}I_{\mathrm{Q,g}}  
\end{array} \right)
=
\left( \begin{array}{rl}
a_{11} & a_{12}\\
a_{21} & a_{22}
\end{array} \right)
\left( \begin{array}{rl}
\mathrm{d}\mu_{\mathrm{g}}\\ 
\mathrm{d}T_{\mathrm{g}}
\end{array} \right),
\end{equation}
where $I_\mathrm{N,g}$ and $ I_\mathrm{Q,g}$ are the particle and heat currents into the gate well.  The `$a$' parameters relate the currents to either a change in chemical potential or temperature assuming the other is constant, e.g. $a_{11} = (\partial I_\mathrm{N,g} / \partial\mu_\mathrm{g})_{\mathrm{d}T_\mathrm{g}=0}$. Equation~(\ref{eqn:Admittance}) is akin to the Onsager relations~\cite{Miller:1959,Onsager:1931} for particle and heat diffusion that have been utilized to describe non-equilibrium transport dynamics in atomic systems~\cite{Brantut:2013he,Hazlett:2013}. Here, they are used to describe the particle and heat currents applied to the gate well in order to elicit the changes $\mu_\mathrm{g} \rightarrow \mu_\mathrm{g} + \mathrm{d}\mu_\mathrm{g}$ and $T_\mathrm{g} \rightarrow T_\mathrm{g} + \mathrm{d}T_\mathrm{g}$.   The particle and heat currents on the left-hand side of Eq.~(\ref{eqn:Admittance}) are given by the differential of the sum of terms in Eqs.~(\ref{eqn:Kirchoff}) and~(\ref{eqn:Energy}), respectively, with current flow into (out of) the gate taken to be positive (negative).  In steady-state, $\mathrm{d}I_\mathrm{Q,g} = 0$; thus, the temperature response to a change in chemical potential is given by $\partial T_\mathrm{g}/\partial \mu_\mathrm{g}=-a_{21}/a_{22}$.  It is not surprising that this quantity is negative, as the chemical potential varies inversely with temperature at constant atom number.  For small deviations from the steady-state, this relation is used in conjunction with Eq.~(\ref{eqn:Admittance}) to determine the input current required to change the gate chemical potential by some $\mathrm{d}\mu_\mathrm{g}$:
\begin{equation}
\mathrm{d} I_\mathrm{g} = \mathrm{d} \mu_\mathrm{g} \left(a_{11}-a_{12}\frac{ a_{21}}{a_{22}}\right),
\end{equation}
where the quantity in the parentheses is equivalent to the gate well input admittance.  The differential current gain about the Q-point is then defined as the ratio of the resulting change in gate-drain current and the gate well input current:
\begin{equation}\label{eqn:CurrentGain}
A_\mathrm{I}\equiv \frac{\mathrm{d} I_\mathrm{gd}}{\mathrm{d} I_\mathrm{g}},
\end{equation}
where $\mathrm{d} I_\mathrm{gd} = I'_\mathrm{gd} - I_\mathrm{gd}$ and the primed current is evaluated at $\mu'_\mathrm{g}=\mu_\mathrm{g}+\mathrm{d}\mu_\mathrm{g}$ and $T'_\mathrm{g}=T_\mathrm{g}+(\partial T_\mathrm{g}/\partial \mu_\mathrm{g}) \mathrm{d}\mu_\mathrm{g}$. 
\begin{figure}[floatfix]
\begin{center}
\includegraphics[width=3.5in]{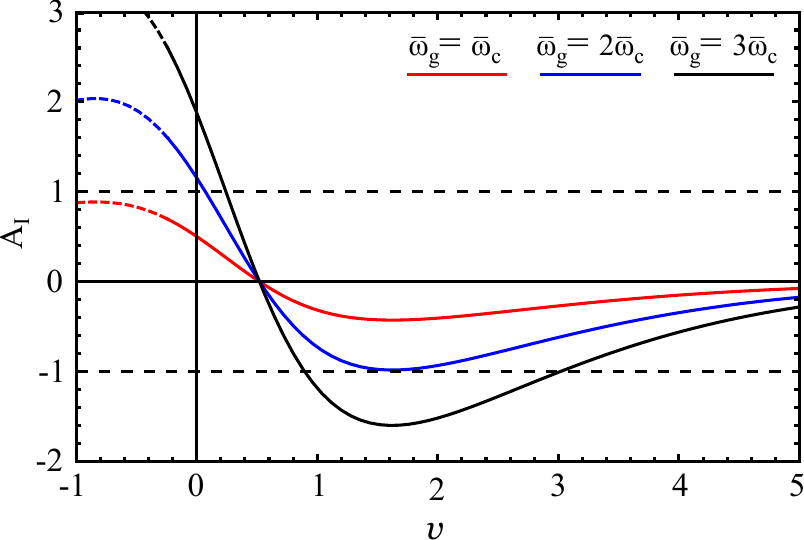}
\caption{\label{fig:CurrentGain} Family of current gain curves in the triple-well system, with $\eta = 5$.  Horizontal dashed lines indicate unity gain. Depending on system parameters, the atomtronic transistor can provide either positive or negative differential current gain. Traces for $\bar{\omega}_\mathrm{g}=1,2$, and $3~\bar{\omega}_\mathrm{c}$ are shown to illustrate the dependence on the mean trap frequency of the gate relative to the external control well.  The traces become dashed where $\upsilon<\upsilon_\mathrm{BEC}$.  The magnitude of the current gain can exceed unity and is controlled by both the gate well trap frequency and feedback parameter. }
\end{center}
\end{figure}

Figure~\ref{fig:CurrentGain} shows the current gain for the range of feedback parameters shown in Fig.~\ref{fig:Temperature_Drop}, illustrating the current gain at various Q-points. Two complementary operating regimes arise in which the transistor provides either positive or negative differential current gain.  For feedback parameters near zero, $A_\mathrm{I}$ is positive, indicating that a positive (negative) $\mathrm{d}I_\mathrm{g}$ leads to an increase (decrease) in $I_\mathrm{gd}$. For increasingly negative feedback parameters, i.e., $\upsilon<\upsilon_\mathrm{BEC}$, the current gain becomes nonphysical as $\mathrm{d}\mu_\mathrm{g}$ becomes ill-defined in the absence of a condensate in the gate well.  At $\upsilon\sim0.5$, the sign of $A_\mathrm{I}$ flips where the changes in $I_\mathrm{gd}$ due to $\mathrm{d}\mu_\mathrm{g}$ and $\mathrm{d}T_\mathrm{g}$ become equal.  Finally, at $\upsilon\sim1.6$, the gain reaches a maximum, negative amplitude.  Positive and negative differential current gain regimes arise due to the interplay between $\mu_\mathrm{g}$ and $T_\mathrm{g}$. To better understand each regime, consider the scaling of the chemical potential with respect to the number of condensed atoms in the Thomas-Fermi limit, $\mu\propto N_\mathrm{c}^{2/5}$. In the negative differential current gain regime, steady-state $\mu_\mathrm{g}$ is larger than in the positive gain regime.  Therefore, a larger number of atoms at $T<T_\mathrm{g}$ must be injected into the gate well to elicit a positive $\mathrm{d}\mu_\mathrm{g}$.  The resulting decrease in $T_\mathrm{g}$ causes a reduction of $I_\mathrm{gd}$ that is more substantial than the increase of $I_\mathrm{gd}$ due to the additional chemical potential. Thus, $\mathrm{d}\mu_\mathrm{g}$ is positive, but the net change in the gate-drain current is negative.  The opposite effect is responsible for positive differential current gain.

With the sign of the current gain understood, the magnitude of the output current is determined.  Multiple traces are shown in Fig.~\ref{fig:CurrentGain} that indicate the gain for different gate well geometric trap frequencies, $\bar{\omega}_\mathrm{g}$, relative to the trap frequencies of the external well that injects the control current, $\bar{\omega}_\mathrm{c}$, where $\bar{\omega}=(\omega_x \omega^2_{\perp})^{1/3}$.  The enhancement in the magnitude of the current gain arises due to the scaling of the specific heat, $C\propto 1/\bar{\omega}^{3}$, and chemical potential, $\mu\propto \bar{\omega}$, of a tight well~\cite{Pethick:2002tn}.  More specifically, as $\bar{\omega}$ increases, the scaling in $\mu$ and $C$ reduce the number of atoms needed to alter the chemical potential and temperature of the steady-state gate well population. In the case that the trap frequencies are equal, one recovers the expected result that $A_\mathrm{I}=1/2$ at $\upsilon = 0$, as there are two equally probable output channels from the gate. Thus, the ratio of trap frequencies is a key design parameter for achieving greater than unity gain.  

\subsection{Power output}\label{sec:ACOp}
\begin{figure}[floatfix]
\begin{center}
\includegraphics[width=3.75in]{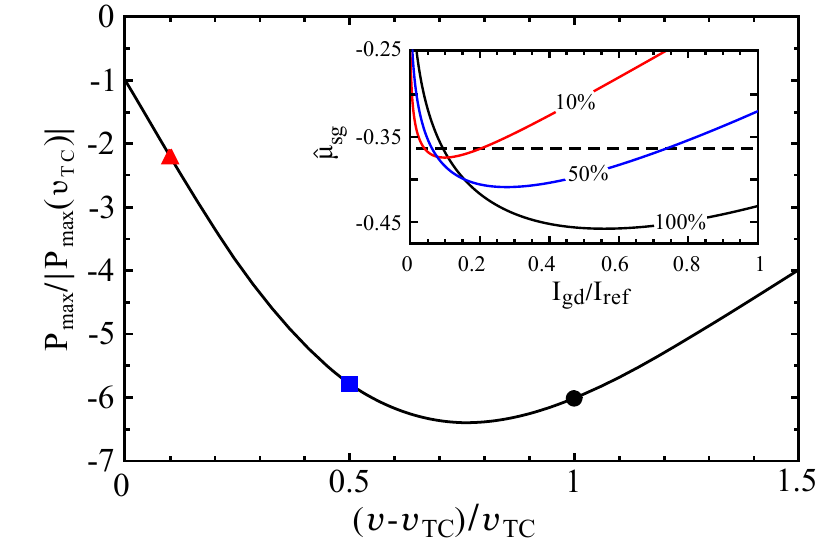}
\caption{\label{fig:Power} Maximum power dissipated when connected to an impedance matched load, normalized to the magnitude of the threshold value, $|P_\mathrm{max}(\upsilon_\mathrm{TC})|$, as a function of the feedback parameter. Above threshold, the transistor exhibits negative transconductance. Inset: Transfer characteristic curves of the atomtronic transistor at $10\%,~50\%$, and $100\%$ above threshold, corresponding to the triangle, square, and circle in the main panel, respectively. Gate-drain current is scaled by a reference current, $I_\mathrm{ref}$, to account for different $\upsilon$. Negative transconductance occurs in regions of negative slope.  The horizontal dashed line indicates the threshold value of $\hat{\mu}_\mathrm{sg}$ that separates positive and negative $g_\mathrm{m}$. }
\end{center}
\end{figure}

Despite a lack of explicit knowledge regarding the load circuit connected to the drain well, it is possible to determine the maximum power delivered to an impedance matched load. Stated generally, impedance matching is the process of selecting the input impedance of the load circuit to be equal to the output impedance of the device such that reflection of the output signal is minimized.  The reflection of particles with a given energy impinging upon a potential landscape can be calculated using a number of methods (e.g.,~\cite{Mokhov:2008cv,Khondker:1988}) and has been experimentally studied using cold~\cite{Cheiney:2013} and ultracold atoms~\cite{Fabre:2011}. For a matched load, the electronics definition for the maximum power at the transistor output~\cite{Chaffee:1933} is:
\begin{equation}\label{eqn:MaxPower}
P_\mathrm{max} = \frac{g_\mathrm{m} \mathrm{d}\mu^2_\mathrm{g}}{4},
\end{equation}
where $\mathrm{d}\mu_\mathrm{g}$ is the amplitude of the gate chemical potential modulation and $g_\mathrm{m}$ is the transconductance, given by
\begin{eqnarray}\label{transconductance}
g_{\mathrm{m}} & \equiv & \frac{\mathrm{d}I_{\mathrm{gd}}}{\mathrm{d}\mu _{\mathrm{sg}}}=\left(\frac{\partial{T_{\mathrm{s}}}}{\partial\mu_{\mathrm{sg}}}\right)\left(\frac{\partial{I_{\mathrm{gd}}}}{\partial{T_{\mathrm{s}}}}\right),
\end{eqnarray}
where the partial derivatives are evaluated at constant $\mu_{\mathrm{s}}$.  Transconductance is useful in describing operation of the devices as an active element, i.e., one that supplies power.  From Eq.~(\ref{eqn:MaxPower}), it can be seen that if $g_m$ is negative, the power dissipated at the transistor output is negative.  Effectively, the transistor converts power supplied by the source well into power output from the drain controlled by $\mathrm{d}\mu_\mathrm{g}$. As alluded to in Fig.~\ref{fig:Temperature_Drop}, there is a threshold feedback parameter, $\upsilon_\mathrm{TC}$, above which the system exhibits negative transconductance in steady-state.  The threshold value is determined from the inflection point of $g_\mathrm{m}$.  The maximum power dissipated at the output of the transistor is shown in Fig.~\ref{fig:Power} as a function of the fraction above threshold, $(\upsilon-\upsilon_\mathrm{TC})/\upsilon_\mathrm{TC}$.  The values of $P_\mathrm{max}$ are calculated using the same fractional value of $\mathrm{d}\mu_\mathrm{g}$ with respect to the steady-state $\mu_\mathrm{g}$ for all $\upsilon$. Both the sign and magnitude of $g_\mathrm{m}$ are determined from the derivative of the transfer characteristic curves.  Examples of these curves are shown in the inset of Fig.~\ref{fig:Power} for $\upsilon = 10\%,~50\%$, and $100\%$ above $\upsilon_\mathrm{TC}$.  The absolute magnitude of the power depends on the value of $\mathrm{d}\mu_\mathrm{g}$; therefore, to better illustrate the behavior of $P_\mathrm{max}$ it is scaled by $P_\mathrm{max}(\upsilon=\upsilon_\mathrm{TC})$ in Fig.~\ref{fig:Power}.  The magnitude of the maximum power dissipated is seen to peak at $\upsilon \approx 1.75~\upsilon_\mathrm{TC} \approx 2.6$.  As was the case for the current gain, the power delivered by the transistor is ultimately limited by the gate-drain current, which decreases exponentially with increasing $\upsilon$.  However, the power is delivered in the form of higher energy thermal atoms.  As a result, the maximum power dissipation peaks at a higher value than the current gain.  

Negative power dissipation, a familiar concept introduced early in the literature on electronic oscillators, originates due to the negative transconductance or transresistance of a device~\cite{Edson:1953vt}.  When coupled to a circuit containing frequency dependent elements, an active device that exhibits negative transresistance cancels the resistive power loss in the load, resulting in the buildup of a resonant, oscillatory signal~\cite{Davis:2011cd,Edson:1953vt}. The oscillator concept is attractive in the field of atomtronics, given the historical impact of RF and higher frequency signal generation within the field of analog electronics.

\section{\label{sec:Expt}Experimental implementation}
The key elements needed to realize the system described in section~\ref{sec:Model} are the trapping potential used to confine the atoms and the barriers that form the triple-well structure.  Overall confinement of the atoms in a cigar shaped potential with $\omega_x \ll \omega_{\perp}$ can be accomplished using either optical or magnetic trapping techniques. Repulsive barriers can be generated by a pair of Gaussian laser beams superimposed on the trap to create the source, gate, and drain wells.  The set of atom currents described in Eq.~(\ref{eqn:currents}) do not rely on tunneling effects, thus the barrier height rather than width is the important design parameter.   To realize the results of section~\ref{sec:Operation}, an additional well must be coupled to the gate. This can be accomplished by  sectioning the source well along the tight trap axis by an additional optical potential.  One half would act as the source well and the other as the control input to the gate.  As an alternative approach, optical trapping techniques have been demonstrated that enable the generation of highly dynamic and reconfigurable trapping potentials~\cite{Pasienski:2008, Henderson:2009}.  Using such techniques, the system could be designed with the control well perpendicular to the direction of source-drain current, ie., a geometry similar to that proposed in Ref.~\cite{Gajdacz:2014}.

A system meeting the criteria of section~\ref{sec:Model} has been used to demonstrate quasi-steady-state forward and reverse biasing of the source-gate junction across a range of Q-points~\cite{Caliga:2015}. In this related work, an atom chip system is used to trap $^{87}$Rb atoms in a magnetic potential with trapping frequencies $(\omega_x, \omega_\perp) = 2\pi\times(67, 1500)~\mathrm{Hz}$. A transparent region of the atom chip in conjunction with a high numerical aperture optical system enables simultaneous in-trap imaging of the atoms and the projection of two blue-detuned optical barriers onto the magnetic trap~\cite{Salim:2013}.  Along the loose axis of the magnetic trap the barriers have a $2.1~\mu\mathrm{m}$ full-width at 1/e and are separated by $4.8~\mu\mathrm{m}$, resulting in a longitudinal gate well trap frequency of $\omega_{x,G} = 2\pi \times 850~\mathrm{Hz}$. During quasi-steady-state operation of the triple-well system, the ratio of source well chemical potential to temperature is $\mu_\mathrm{s}/T_\mathrm{s} \gtrsim 0.4$ and the truncation parameter is $\eta \gtrsim 3$.  

\section{Conclusion}
Finite temperature analogues to electronic components represent a complementary approach to those that utilize the superfluid behavior of ultracold Bose gases.  Ultimately, the effects of dissipation and entropy transfer associated with transistor action and information processing must be accounted for.  Rather than starting from a fully quantum system and implementing dissipation as a perturbation, we presented a semiclassical model of a tailored triple-well system that utilizes the interplay of thermodynamic variables to exhibit transistor action.

Within this approach, atom currents driven by chemical potentials and thermal energy are used to describe the particle and energy transport throughout the system.  By varying the height of the barriers that separate the three wells, it was shown possible to control the steady-state chemical potential and temperature differences between the source and gate wells, an analogous process to biasing an electronic transistor to the desired quiescent point.  Within the range of Q-points that were explored, the power dissipation within the transistor was discussed.  Regimes of both positive and negative static resistance were found, indicating either net heat dissipation or extraction from within the transistor.    

Transistor-like behavior of the triple-well atomic system was investigated by calculating the current gain given an external current input to the gate well. The device exhibits regions of both negative and positive differential current gain, depending on the feedback parameter, that can exceed unity.  The magnitude of the gain was shown to depend on the mean trap frequency of the gate well, a characteristic that arises due to the nonlinear dependence of the chemical potential and specific heat on the trap frequency. Therefore, the current gain is a widely tunable parameter via both $\upsilon$ and $\bar{\omega}_\mathrm{g}$.

In addition to transistor action, the maximum power delivered to an impedance matched load was calculated. Again, given an external current input to the gate well, the transistor exhibits negative power dissipation at its output.  Thus, the triple-well potential supplies power to a load circuit.  As in analog electronics, active devices, i.e., those which exhibit negative transresistance behavior, can be coupled with frequency dependent loads to create an oscillatory output signal.  Therefore, if coupled to the appropriate `load,' the transistor-like triple-well system described here could be used to generate oscillating atomic currents.

\ack
This work was supported by the Air Force Office of Scientific Research (FA9550-14-1-0327), the National Science Foundation (PHY1125844), and by the Charles Stark Draper Laboratories (SC001-0000000759).

\end{document}